\documentclass[conference]{IEEEtran}
\IEEEoverridecommandlockouts

\usepackage{cite}

\usepackage{amsmath,amssymb,amsfonts}
\usepackage{algorithmic}
\usepackage{graphicx}
\usepackage{textcomp}
\usepackage{subcaption}
\usepackage[acronym]{glossaries}
\makeglossaries
\newacronym{ran}{RAN}{Radio Access Network}
\newacronym{oran}{O-RAN}{Open Radio Access Network}
\newacronym{qos}{QoS}{Quality of Service}
\newacronym{qoe}{QoE}{Quality of Experience}
\newacronym{ott}{OTT}{Over-The-Top}
\newacronym{ibn}{IBN}{Intent-Based Networking}
\newacronym{llm}{LLM}{Large Language Model}
\newacronym{llms}{LLMs}{Large Language Models}
\newacronym{xapp}{xApp}{extended application}
\newacronym{ric}{RIC}{RAN Intelligent Controller}
\newacronym{nrt-ric}{near-RT RIC}{near-real-time RAN Intelligent Controller}
\newacronym{e2}{E2}{E2 interface}
\newacronym{kpi}{KPI}{Key Performance Indicator}
\newacronym{kpis}{KPIs}{Key Performance Indicators}
\newacronym{kpm}{KPM}{Key Performance Measurement}
\newacronym{tc}{TC}{Traffic Control}
\newacronym{tc-ran}{TC-RAN}{Traffic Control Radio Access Network}
\newacronym{oai}{OAI}{OpenAirInterface}
\newacronym{ue}{UE}{User Equipment}
\newacronym{rlc}{RLC}{Radio Link Control}
\newacronym{drll}{DRL}{Deep Reinforcement Learning}
\newacronym{ml}{ML}{Machine Learning}
\newacronym{sla}{SLA}{Service Level Agreement}
\newacronym{prb}{PRB}{Physical Resource Block}
\newacronym{rtt}{RTT}{Round-Trip Time}
\newacronym{isr}{ISR}{Intent Satisfaction Ratio}
\newacronym{codel}{CoDel}{Controlled Delay}
\newacronym{rr}{RR}{Round Robin}
\newacronym{udp}{UDP}{User Datagram Protocol}
\newacronym{tcp}{TCP}{Transmission Control Protocol}
\newacronym{json}{JSON}{JavaScript Object Notation}
\newacronym{aqm}{AQM}{Active Queue Management}
\newacronym{osi}{OSI}{Open Systems Interconnection}
\newacronym{cqi}{CQI}{Channel Quality Indicator}
\newacronym{mcs}{MCS}{Modulation Coding Scheme}
\newacronym{cca}{CCA}{Congestion Control Algorithm}
\usepackage{svg}
\usepackage{balance}
\usepackage{xcolor}
\usepackage{booktabs}
\usepackage{multirow}
\usepackage{url}
\usepackage{comment}

\def\BibTeX{{\rm B\kern-.05em{\sc i\kern-.025em b}\kern-.08em
    T\kern-.1667em\lower.7ex\hbox{E}\kern-.125emX}}

\begin{document}
\bstctlcite{IEEEexample:BSTcontrol}
\title{An LLM-Driven Traffic-Control xApp \\for Dynamic 5G RAN Intents\\
\thanks{This work is supported by the Chips Joint Undertaking (JU), European Union (EU) HORIZON-JU-IA, under grant agreement No. 101140087 (SMARTY). Results presented in this paper were obtained using the Chameleon testbed supported by the National Science Foundation.}
}

\author{
    \IEEEauthorblockN{Valerio Triolo$^{1}$, Luca Valcarenghi$^{1}$, and Emilio Paolini$^{1}$}
    \IEEEauthorblockA{$^1$\textit{TeCIP Institute, Scuola Superiore Sant'Anna, Pisa, Italy}}  
    \IEEEauthorblockA{valerio.triolo@santannapisa.it}
}

\maketitle

\begin{abstract}
Programmable Radio Access Networks (RANs) enable runtime deployment of traffic-control functions, but selecting the appropriate configuration remains challenging as network conditions and service requirements change over time. Conventional solutions often rely on static rules or learning-based controllers with predefined state and action spaces, limiting their ability to accommodate changing intents. This paper introduces an LLM-driven xApp for traffic control in programmable RANs. The proposed architecture combines operator-defined service intents, runtime E2 interface telemetry, and the currently deployed configuration to generate traffic-management policies without blocking packet forwarding or the processing of incoming E2 measurements during LLM inference. Experiments on an OpenAirInterface 5G environment show that the proposed controller dynamically adapts the traffic-control pipeline to runtime intent changes, achieving an average intent satisfaction ratio of $0.91$, compared with $0.52$ for a static TC-RAN configuration. When channel conditions degrade, the controller maintains an intent satisfaction ratio of $\approx 0.88$, compared with $\approx 0.4$ for the static baseline.

\end{abstract}

\section{Introduction}
Current 5G and beyond-5G networks must support heterogeneous applications~\cite{NextGeneration5G}, ranging from video streaming and interactive gaming to real-time communication and emerging AI-driven services. The \gls{qoe} of these applications may depend on both application-level factors, such as encoding bitrate, resolution, and playout buffering, and network-level \glspl{kpi}, such as throughput, latency, jitter, and packet loss. To assess service quality, customer satisfaction, and performance degradation, \gls{ott} content providers monitor application-level \gls{qoe} through dedicated models~\cite{perceptual}. When this feedback is shared with the network operator, it can help identify the network requirements needed to improve service quality. These requirements can be expressed as network-level \emph{intents}, i.e., declarative, high-level descriptions of the desired network behavior, complemented by quantitative constraints on network \glspl{kpi}, such as minimum throughput, maximum latency, or maximum packet loss~\cite{leivadeas2022survey}. As QoE feedback from OTT providers changes, the operator may update the active intent accordingly, for example, prioritizing lower packet loss over throughput to improve service reliability. However, translating an intent, including both its high-level objectives and its constraints, into an appropriate traffic-control configuration is not straightforward: traffic-control decisions involve trade-offs between \glspl{kpi} when multiple services compete for shared resources, and a configuration that satisfies an intent under given conditions may become ineffective when the intent or the network conditions change.

Programmable \gls{ran} architectures expose traffic-control mechanisms, such as packet classification, queue management, pacing, and traffic shaping, which authorized control applications can reconfigure within the permissions granted by the underlying framework. For example, frameworks such as TC-RAN~\cite{irazabal2023tc} enable reconfigurations at runtime. However, programmability addresses how a selected policy is enforced, rather than how the most appropriate policy is selected~\cite{leivadeas2022survey}. In TC-RAN, the logic that selects the pipeline is defined at design time, which limits its flexibility when the active intent or the network conditions change. Selecting a suitable pipeline and its parameters is itself challenging, as it requires jointly considering the active intent, the current traffic composition, and the observed network state~\cite{Bonati2021}.

In this direction, \glspl{llm} are emerging as a promising tool to introduce higher-level reasoning and \gls{ibn} principles into network control applications~\cite{LLMenabled}. When integrated into a \gls{ran} intelligent controller (RIC), e.g., via an xApp, an LLM can process high-level service requirements together with runtime network measurements and translate them into appropriate control decisions. 
However, directly using an \gls{llm} in a network control loop raises several challenges~\cite{wu2024netllm}. LLM outputs may be inconsistent, syntactically invalid, or incompatible with the current network state~\cite{mondal2023llms}. In addition, the latency and reliability of the decision process must be taken into account before a generated configuration can be safely applied~\cite{HashemiNatanzi2026AdvancedAS}. Therefore, an LLM-based controller requires a limited action space, structured input and output, and a validation stage that prevents unsafe or unsupported configurations from reaching the network.

To address the above challenges, we propose an \gls{llm}-driven xApp for traffic control in programmable RAN environments. The main contributions of this work are threefold. First, we design an intent-driven control architecture that combines high-level service intents, runtime E2 telemetry, and the currently deployed configuration to perform \gls{llm}-based policy selection over programmable traffic-control mechanisms. Second, we introduce a constrained policy-generation workflow based on a predefined action space, structured outputs, and deterministic validation before enforcement. Third, we experimentally demonstrate closed-loop adaptation to changing multi-\gls{kpi} service intents and channel conditions on an OpenAirInterface-based 5G platform, evaluating the approach across different \gls{llm} backends and different combinations of intent thresholds and priorities. The proposed controller achieves an average intent satisfaction ratio of 0.91, compared with 0.52 for a static TC-RAN configuration, while also maintaining a high intent satisfaction ratio under degraded radio channel conditions.

\section{Related work}
The programmability introduced by Open \gls{ran} enables third-party applications to monitor network conditions and dynamically control selected \gls{ran} functions. In particular, xApps running on the near-real-time \gls{ran} Intelligent Controller (near-RT RIC) have been investigated for traffic steering, resource allocation, mobility management, and network slicing~\cite{Marinova2024}. Within this context, TC-RAN \cite{irazabal2023tc} extends the O-RAN control framework toward programmable traffic control by exposing traffic-management mechanisms to near-RT RIC applications. This enables xApps to dynamically configure different functions, such as classification, queue management, pacing, and shaping, providing a programmable control layer between \gls{ran}-level decisions and end-host traffic control. However, while TC-RAN makes these mechanisms configurable, it does not decide how they should be configured. The pipeline enforced by a TC-RAN xApp, i.e., which components are installed and with which parameters, is fixed when the xApp is designed. When the operator issues a new intent whose requirements the deployed pipeline cannot satisfy, or the network reaches another state, the xApp must be stopped, a pipeline suited to the new requirements must be identified, and the xApp must be redeployed. Until this cycle completes, the network keeps operating with a configuration that is misaligned with the active intent. Moreover, identifying an appropriate pipeline that satisfies a given intent is a non-trivial task for a human: components and parameters affect latency, loss, and throughput jointly, often with opposite effects. This motivates controllers that automatically select the configuration matching the active requirements, rather than relying on the operator to do so.

A first class of such controllers relies on \gls{ml}, which is increasingly used in \gls{qos} together with \gls{sla} management, to dynamically optimize network resources. Recent studies have proposed \gls{drll} approaches for robust QoS management within RAN slicing. For example, RSLAQ~\cite{yungaicela2026rslaq} implements an xApp that maps operator-defined \gls{sla} requirements into radio-resource control decisions, using \gls{drll} to adapt \gls{prb} allocation and scheduling across network slices. However, the controller operates over a predefined state, action, and reward space tailored to the considered slicing scenario. As a result, accommodating new service objectives or control mechanisms may require redesigning the learning formulation and retraining the model.

At the intent-translation layer, Intent2QoS~\cite{intent2qos} uses few-shot prompting and a deterministic critic to convert traffic-shaping intents into Linux traffic-control configurations. However, it is evaluated offline, by similarity to expert-defined references on a synthetic dataset, and does not consider live telemetry, runtime intent changes, or the QoS achieved after enforcement.

A closely related approach is \gls{llm}-xApp~\cite{wu2025llmxapp}, which uses an \gls{llm} as an online optimizer for radio resource management in O-RAN: at each step, a meta-prompt with past observations, requested data rates, and evaluation scores yields new per-slice \gls{prb} allocations, enforced through E2 control messages on the OAIC platform. While it demonstrates the feasibility of an LLM in an operational O-RAN control loop, its action space is limited to inter-slice radio resources and its requirements are expressed mainly as data rates. It therefore does not address multi-\gls{kpi} intents nor the composition of packet-processing mechanisms such as queue management, classification, pacing, and shaping.

In contrast to prior work, our objective is not to use the LLM as a direct resource allocator or as an offline configuration generator. Instead, we investigate whether an LLM can act as a runtime policy-selection layer over a programmable traffic-control substrate, jointly considering service intents, live network telemetry, and the currently deployed configuration. 

\section{System Architecture}
\begin{figure}[!ht]
    \centering
    \includegraphics[width=1\columnwidth]{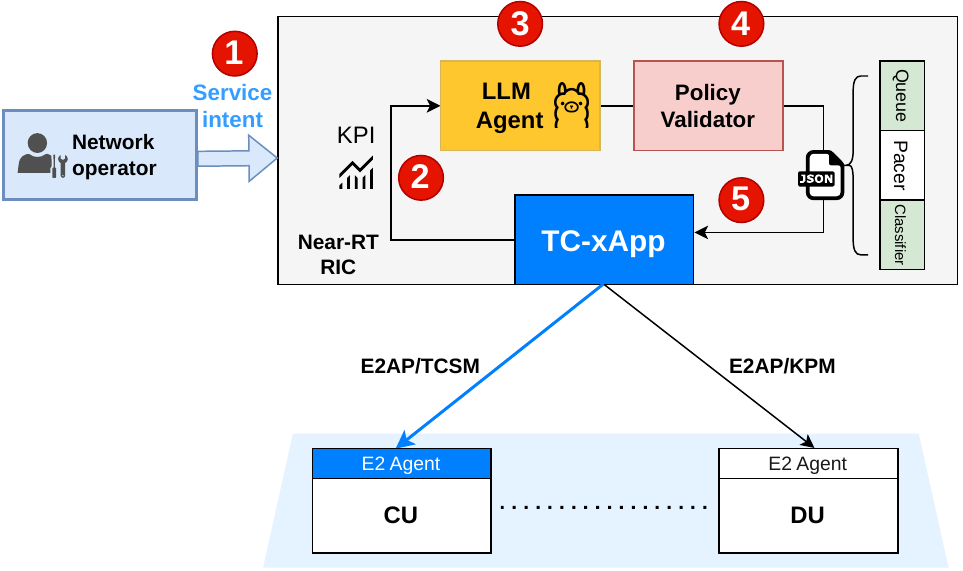}
    \caption{Architecture and closed-loop workflow of the proposed LLM-driven xApp. Runtime telemetry and service intents are processed by the LLM agent to generate a TC-RAN configuration, which is enforced at the gNB through the E2 interface.}
    \label{fig:architecture}
\end{figure}

In conventional programmable \gls{ran} control, traffic-management mechanisms can be dynamically configured, but the logic to select the appropriate policy is typically predefined. However, as service requirements or network conditions change, a previously deployed configuration may become ineffective. Our system introduces an \gls{llm}-based policy-selection mechanism into the near-RT O-RAN control loop. As shown in Fig.~\ref{fig:architecture}, the architecture comprises a RAN node, an xApp running on the near-RT RIC, an LLM-based policy agent, and a Policy Validator. The network operator submits the intent to the xApp (1). The LLM then generates a policy based on the intent, the current KPI measurements (2), received through the E2 interface, and the system prompt (3). After validation (4), the resulting configuration is passed to the TC-xApp, which translates it into the corresponding control messages (5) and dispatches them to the RAN node again through the E2 interface. 


\subsection{LLM-based Policy Selection}
A service intent, submitted by the network operator, specifies the desired operational objectives and their associated constraints. In our implementation, an intent defines bounds on different \gls{kpi}s, such as latency, UDP packet loss, and throughput. At each decision instant, the xApp combines the active intent, latest E2 telemetry, and deployed configuration into a structured prompt. A fixed system prompt defines the available TC-RAN actions and output schema, while the user prompt provides the current network context. This formulation lets new \gls{kpi}s, qualitative operator priorities, or additional TC-RAN components be introduced by extending the system prompt and the validator itself, without redesigning hand-crafted rules. The decision-making process is triggered both periodically and upon a change in the active service intent. To ensure uninterrupted data-plane operations, the xApp captures a network state snapshot and invokes the LLM asynchronously. Thus, the LLM inference latency does not block packet forwarding or processing of incoming E2 measurements. Furthermore, if multiple triggers occur while an inference is pending, the requests are discarded to prevent redundant LLM calls.

Upon completion, the \gls{llm} outputs a high-level policy formatted as a JSON object. The model can either propose a new processing pipeline or recommend maintaining the current configuration when no reconfiguration is necessary. If an update is required, the policy specifies an ordered list of traffic-control components and their associated parameters. The available components are: an OSI-based traffic classifier, which maps incoming packets to separate queues based on header fields at different layers; a \gls{codel} queue~\cite{codel}, which drops packets when their queuing delay persistently exceeds a target, thereby bounding latency; a scheduler, which determines the order in which queues are served; a per-queue traffic shaper, which limits the rate of a given queue; and a 5G-BDP pacer, which limits the data forwarded to the \gls{rlc} buffer to the estimated bandwidth-delay product of the radio link.

\subsection{Policy Validation and Enforcement}
Since \gls{llm}-generated outputs may be syntactically invalid or inconsistent with the operational context, the generated policy is never applied directly. Instead, it passes through a deterministic validation stage that establishes a safety boundary between the LLM output and the executable traffic-control configuration. First, the output is parsed and evaluated against an expected JSON schema to verify that all required fields, data types, and policy structures are valid. Next, the validator cross-references the proposed actions against a whitelist of admissible operations, iterating over the pipeline components and discarding any that fall outside the supported TC-RAN action space. To prevent the LLM from inadvertently degrading network performance, all numerical parameters are subject to boundary clipping. For example, the \gls{codel} maximum acceptable queuing delay (\texttt{target\_ms}) is clamped between $5$ and $100$ milliseconds. Moreover, the validator provides a fallback mechanism for invalid or unsupported configurations: in the event of an invalid or misconfigured pipeline, it automatically rolls back to a default configuration, which consists of deploying a \gls{codel} queue alongside an OSI classifier for TCP isolation. 

Finally, the validated policy is translated into a sequence of control messages, and they are dispatched to the \gls{ran} node via the E2 interface.

\section{Performance evaluation}

\begin{figure}[t]
    \centering
    \includegraphics[width=\columnwidth]{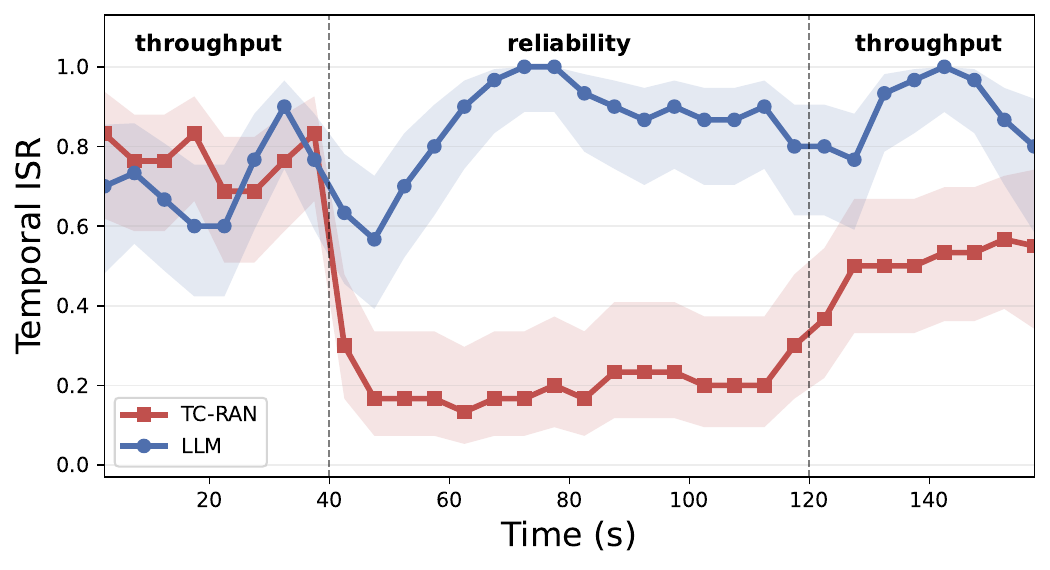}
    \caption{Temporal evolution of the Intent Satisfaction Ratio (ISR)
    under runtime intent changes, following a throughput--reliability--throughput sequence.}
    \label{fig:temporal_isr}
\end{figure}

\subsection{Experimental Setup and Metrics}

\begin{figure*}[t]
    \centering
    \begin{subfigure}{0.32\textwidth}
        \includegraphics[width=\linewidth]{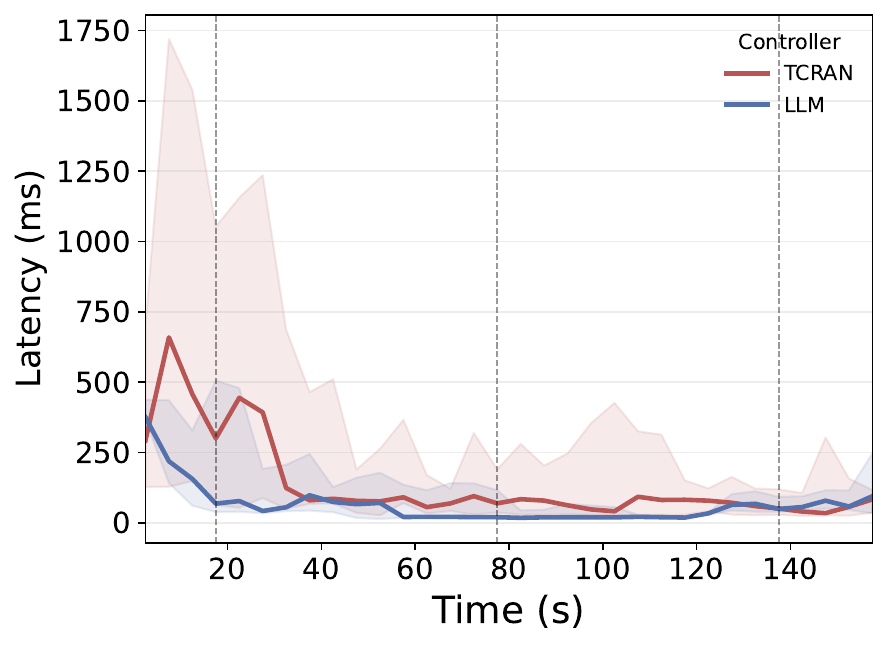}
        \caption{95th-percentile UDP latency.}
        \label{fig:kpi_latency}
    \end{subfigure}\hfill
    \begin{subfigure}{0.32\textwidth}
        \includegraphics[width=\linewidth]{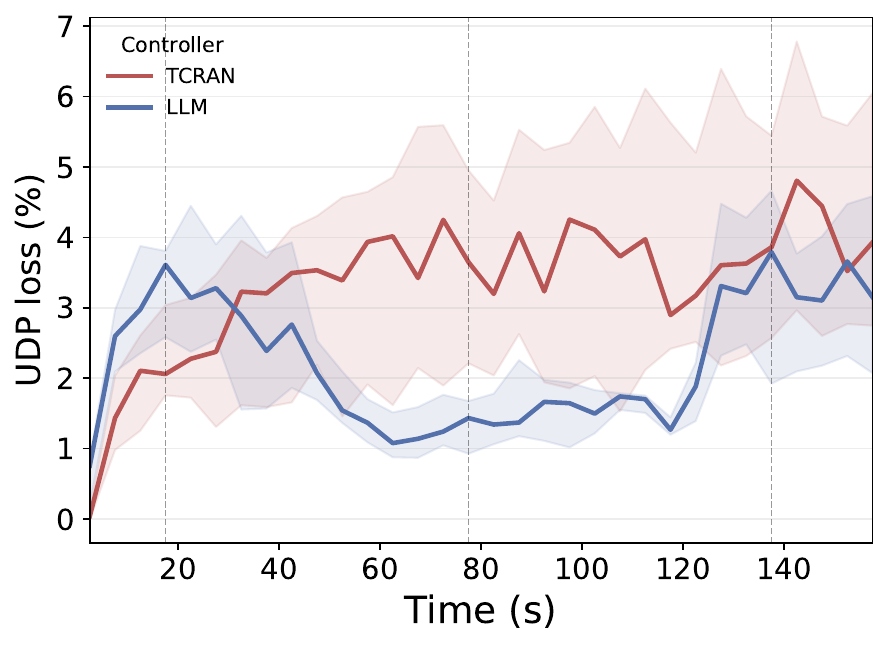}
        \caption{UDP packet loss ratio.}
        \label{fig:kpi_loss}
    \end{subfigure}\hfill
    \begin{subfigure}{0.32\textwidth}
        \includegraphics[width=\linewidth]{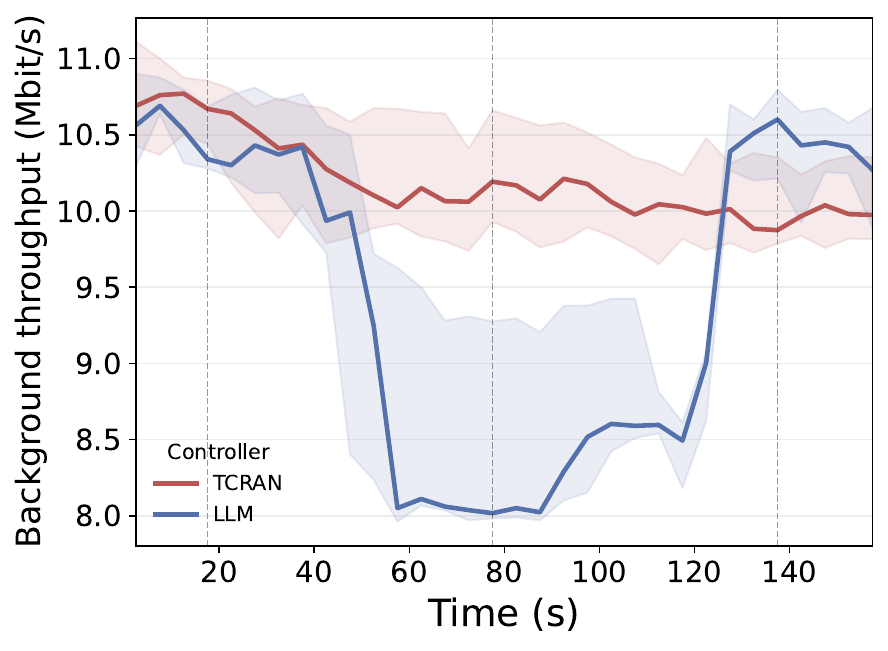}
        \caption{Background TCP throughput.}
        \label{fig:kpi_throughput}
    \end{subfigure}
    
    \caption{Temporal evaluation of the key performance indicators (latency, packet loss, and throughput) under the LLM-driven controller and the static TC-RAN baseline.}
    \label{fig:kpis}
\end{figure*}

We prototyped the \gls{llm}-powered \gls{tc} xApp on top of the OAI 5G SA stack from the official repository\footnote{https://gitlab.eurecom.fr/oai/openairinterface5g/} \cite{oai}. The testbed comprises an OAI gNB (commit \texttt{48cce17c}), one OAI UE (commit \texttt{50d9c1f}), a 5G Core (commit \texttt{1c2f4d3c}), and FlexRIC (commit \texttt{7ff45e8a}). Radio access was emulated through the OAI RF simulator using NR band~$78$, numerology~1, and $106$ physical resource blocks. The near-RT RIC and the TC-RAN xApps were implemented using FlexRIC, while the gNB exposed RLC, PDCP, traffic-control measurements and operations through the E2 interface.


All experiments consider downlink traffic from an application server, located inside the Data Network, to the UE, and generated with \textit{iperf3}. The workload combines two concurrent flows on distinct transport-layer ports, allowing the TC-RAN classifier to separate them. A UDP flow emulates a time-sensitive service, e.g., a real-time voice or gaming service, with a constant bit rate of $1$~Mbit/s and 160-byte payloads. The background load consists of six parallel TCP streams using CUBIC as \gls{cca}, which saturate the radio link and fill the \gls{rlc} buffer. In each run, UDP and TCP traffic start at the same instant and coexist for $160$~s, during which the active intent changes according to a fixed schedule. Each configuration is repeated $10$ times under identical conditions, and the reported 95\% confidence intervals are computed from the results of these independent runs.

We consider two control approaches. The \emph{static TC-RAN} baseline deploys a predefined traffic-control pipeline when the RLC sojourn
time first exceeds $10$~ms. This configuration consists of a CoDel queue with default parameters, an OSI classifier, and a 5G-BDP pacer, as adopted in~\cite{irazabal2023tc}. The proposed \emph{LLM-driven TC-RAN} approach operates under the same traffic and network conditions, but receives a notification whenever the active service intent changes. Upon notification, it can either preserve the current pipeline or generate a new configuration including a 5G-BDP pacer, a scheduler, a CoDel queue, traffic classifiers, and per-queue shaping parameters. 

During each run, the active service intent changed from a throughput-oriented requirement to a reliability-oriented requirement and then back to the throughput-oriented requirement. Performance was evaluated over non-overlapping windows of $\Delta=5$~s, computing for each window $k$ and repetition $r$ the $95$th-percentile UDP latency $L_{r,k}^{95}$, the packet-weighted UDP loss ratio $P_{r,k}$ (lost over transmitted packets in the window), and the average aggregate background TCP throughput $B_{r,k}$.

Because the system must satisfy all constraints associated with the active service intent simultaneously, we model the outcome using a binary intent-satisfaction indicator. Let $i(k)$ denote the active intent during window $k$, and let $\mathcal{C}_{i(k)}$ be the set of performance constraints imposed by that intent. Let $m_{r,k}^{(c)}$ represent the specific metric measured for constraint $c \in \mathcal{C}_{i(k)}$ during repetition $r$ and window $k$. The binary intent satisfaction $S_{r,k}$ is defined as:
$
S_{r,k} =
\mathbb{I}\left[
\bigwedge_{c \in \mathcal{C}_{i(k)}}
\left(m_{r,k}^{(c)} \unlhd_c \tau_c\right)
\right],
$
where $\tau_c$ is the threshold specified by constraint $c$, and $\unlhd_c$ is the relational operator associated with that constraint: $\unlhd_c$ is $\leq$ for upper-bound constraints and $\geq$ for lower-bound constraints. Each comparison $m^{(c)}_{r,k} \unlhd_c \tau_c$ is therefore true when constraint $c$ is met. The operator $\bigwedge$ denotes the logical AND over all constraints of the active intent, and $\mathbb{I}[\cdot]$ is the indicator function, which returns 1 if its argument is true and 0 otherwise. Hence, $S_{r,k}=1$ only when all requirements imposed by the active intent are simultaneously satisfied in window $k$, and $S_{r,k}=0$ otherwise.


To quantify the effectiveness of the controller in maintaining the desired operational state over time, we use the \gls{isr} as the primary evaluation metric. For a given time window $k$, the temporal \gls{isr} is computed across the $N_k$ repetitions as
$
\mathrm{ISR}(k) =
\frac{1}{N_k}
\sum_{r=1}^{N_k} S_{r,k},
$
where \(N_k\) denotes the number of independent runs considered for a time window $k$. 
In each run, the active intent is changed at the same predefined instants, and every window $k$ refers to the same intent across all repetitions.
For the policy generation, local \gls{llm} inference was served via \texttt{Ollama} on a remote machine equipped with an NVIDIA Tesla P100 GPU and 16~GB of VRAM. All experiments using the \gls{llm}-based controller employ Qwen3-8B with a context window of $3072$~tokens, a temperature of $0$, a maximum generation limit of $256$~output tokens and the thinking mode of the model disabled. On this setup, a policy decision with Qwen3-8B requires an average inference time of $1.47$~s, while installing the resulting pipeline at the gNB takes $1.22$~ms.

\begin{figure}[t]
    \centering
    \includegraphics[width=1\columnwidth]{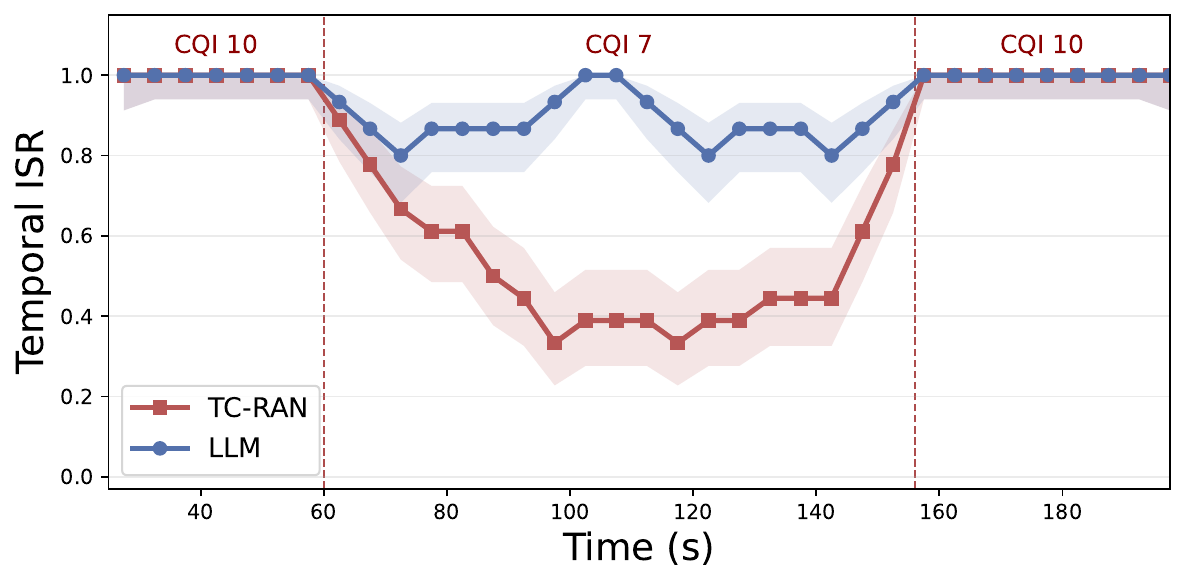}
    \caption{Temporal \gls{isr} under a channel-quality degradation with a
    fixed service intent: the equivalent \gls{cqi} at UE drops from 10 to 7 at $t=60$~s and is
    restored at $t\approx155$~s.}
    \label{fig:channel}
\end{figure}

\subsection{Runtime Intent Adaptation}

We first evaluate the ability of the proposed controller to adapt the traffic-control configuration when the active service intent changes. 
Fig.~\ref{fig:temporal_isr} reports the temporal \gls{isr} and the corresponding confidence intervals obtained by the static TC-RAN baseline and the proposed LLM-driven controller under the throughput--reliability--throughput sequence averaged over $10$ runs.
During the initial throughput-oriented phase, both controllers operate mostly within the intent constraints. In the reliability-oriented phase, the LLM-driven controller maintains an ISR mostly between $0.8$ and $1.0$, while the static baseline remains around $0.2$. 
The drops observed for the LLM-driven controller, specifically at time instants $30$ and $120$, are mainly caused by reconfiguration transients. Indeed, after a new policy is enforced, the KPIs do not change instantaneously: queued traffic must first drain, TCP flows have to adapt to the new rate, and the TCP congestion control needs several round-trip times to converge to the new shaping or pacing rate, leading to short-time constraint violations.
To identify the causes of intent violations, in Fig.~\ref{fig:kpis} we report the evolution of the main performance indicators, namely latency, UDP packet loss, and background TCP throughput.
\subsection{Adaptation to Channel Variations}
The previous experiment considered intent changes under stable network
conditions. However, intents typically change on long time scales, whereas network indicators, such as the radio \gls{cqi}, may vary continuously. To assess whether the controller can react to such variations, Fig.~\ref{fig:channel} reports the temporal \gls{isr} of the static TC-RAN baseline and of the LLM-driven controller when the throughput-oriented intent is kept fixed for the whole run, which lasts $190$~s, while the radio channel is degraded at runtime. To emulate a time-varying channel, we replay a downlink \gls{mcs} trace at the gNB MAC scheduler: the \gls{mcs} is set to $16$ until $t=60$~s, reduced to $9$ until $t\approx155$~s, and then restored; this reduces the link capacity from approximately $24$ to $11.5$~Mbit/s. Since \gls{mcs} selection is closely related to channel quality, Fig.~\ref{fig:channel} reports the equivalent \gls{cqi} values rather than \gls{mcs} indices, as CQI provides a more direct indication of the radio conditions experienced by the UE.


Under nominal channel conditions, i.e., \gls{cqi}=10, both controllers satisfy the intent. However, when the channel quality degrades, i.e., \gls{cqi}=7, the static TC-RAN configuration cannot adapt to the reduced link capacity, allowing TCP traffic to saturate the radio link and causing the \gls{isr} to drop to $\approx 0.35$--$0.45$ throughout the degraded phase. In contrast, the LLM-driven controller dynamically introduces a traffic shaper on the TCP queue, limiting its rate to $\approx 11$~Mbit/s. By regulating the TCP traffic, the controller reduces buffer occupancy and limits contention with the latency-sensitive UDP flow, improving packet delivery while preserving the required TCP throughput. As a result, it achieves an average \gls{isr} of $\approx 0.88$ under degraded channel conditions.

\begin{figure}[!ht]
\centering
\includegraphics[width=1\columnwidth]{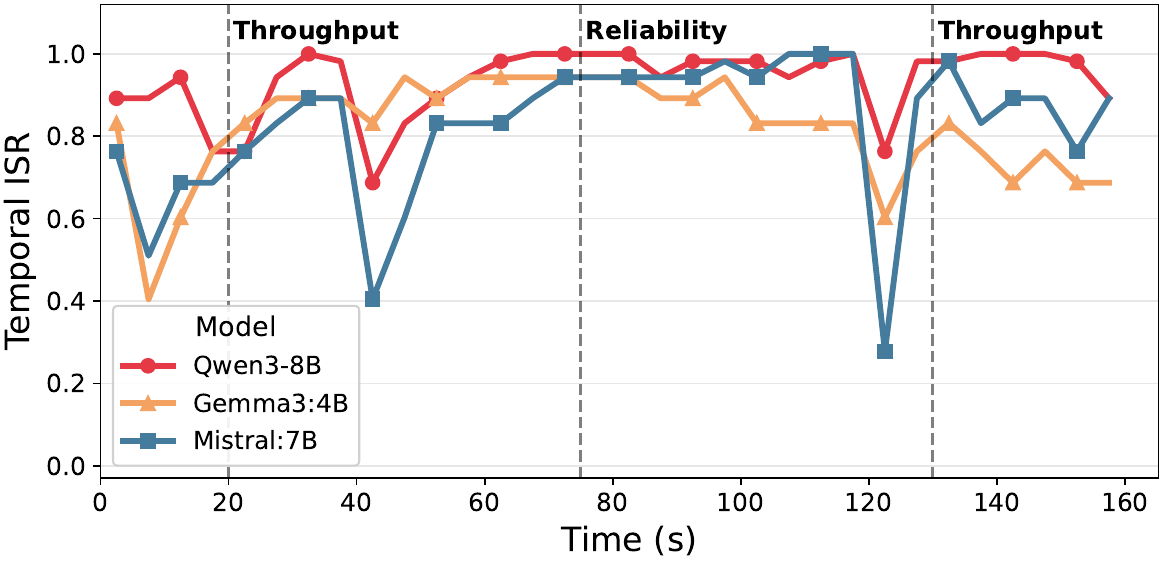}
\caption{Temporal \gls{isr} obtained with different LLMs under the throughput--reliability--throughput intent sequence. Vertical dashed lines denote intent transitions.}
\label{fig:llm-comparison}
\end{figure}

\subsection{Impact of the LLM Backend}

To assess whether the control performance is sensitive to the choice of the \gls{llm}, we ran a separate experimental campaign  comparing Qwen3-8B with Mistral-7B and Gemma3-4B under the same traffic conditions, intent sequence, and control architecture.

As shown in Fig.~\ref{fig:llm-comparison}, Qwen3-8B achieves the highest average \gls{isr} ($\approx 0.91$), followed by Mistral-7B ($\approx 0.83$) and Gemma3-4B ($\approx 0.82$).
The gap is mainly due to the throughput-oriented phases: after the initial transient, Qwen3-8B remains close to one, whereas Mistral-7B and Gemma3-4B show larger fluctuations. In the reliability-oriented phase, all three models satisfy the active intent for most of the interval.

Across all models, between $98.6\%$ and $100\%$ of the generated policies passed validation, indicating that the restricted action space and the deterministic validator allow the architecture to operate reliably regardless of the underlying \gls{llm}. The residual differences in \gls{isr} may also depend on inference latency and the time required to apply updated configurations, and should therefore not be read solely as differences in reasoning capability. Optimizing inference latency (e.g., through quantization or smaller models) is outside the scope of this work, which focuses on the feasibility of LLM-driven policy selection.

\subsection{Discussion and Scalability}

The main motivation of this work is the ability to accommodate service intents with different KPI thresholds and priorities without relying on predefined mappings between intents and configurations. Rule-based controllers require explicit logic to map service requirements and network conditions to control actions, while profile-based approaches rely on predefined configurations. Both approaches may become increasingly difficult to maintain as the number of requirements and their combinations grows.


For instance, a profile-based controller relies on predefined configurations to cover the combinations of control parameters and traffic queues anticipated at design time, so the number of possible configurations can grow rapidly with the size of the configuration space and choosing the correct configuration can be time-consuming. 
The proposed LLM-driven controller addresses this challenge by interpreting the active intent together with runtime telemetry and the currently deployed configuration. Based on this information, the LLM selects a traffic-control policy and specifies its parameters within the action space supported by the xApp. Consequently, different combinations of KPI thresholds and priorities can potentially be handled without defining a dedicated rule or profile for each combination, by simply submitting an intent. 

To assess this capability, we evaluate the controller using five additional, previously unseen service intents. Each intent specifies constraints on three KPIs: 95th-percentile UDP latency, UDP packet loss, and minimum background TCP throughput. Their respective thresholds are generated using Latin hypercube sampling over $[180,450]$~ms, $[1.5,5.0]\%$, and $[5.0,9.5]$~Mbit/s, with seed \texttt{20260917}. Each intent is also labeled with a primary priority—latency, reliability, or throughput—which is communicated to the LLM as a qualitative preference when selecting a policy. The five intents are evaluated over $10$ independent runs, with their execution order randomized across repetitions. This experiment tests generalization to new combinations of KPI thresholds and priorities.

\begin{figure}[t]
    \centering
    \includegraphics[width=1\columnwidth]{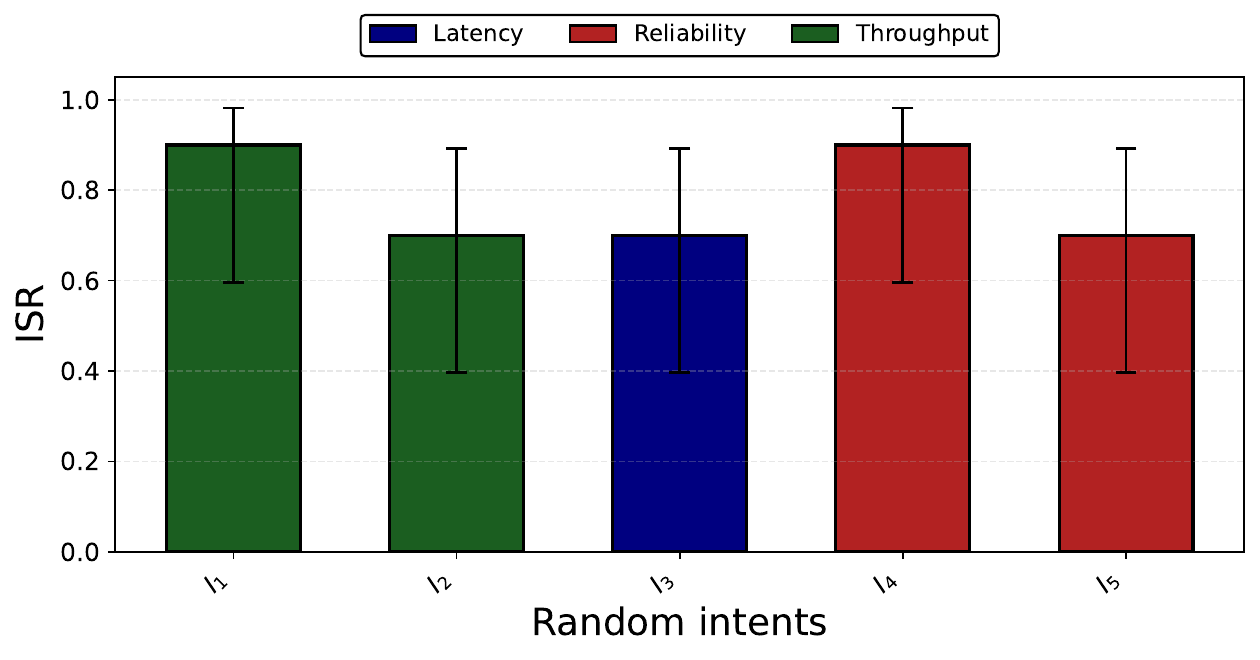}
    \caption{\gls{isr} of the LLM-driven xApp on five held-out intents
    ($I_1, \dots, I_5$) combining the same three \glspl{kpi} with unseen
    threshold values and priorities. Bar colors indicate the primary priority
    of each intent.}
    \label{fig:generalization}
\end{figure}

Fig.~\ref{fig:generalization} reports the \gls{isr} for each intent, averaged over the $10$ runs, with 95\% confidence intervals. Across the evaluated runs, the controller deploys five distinct configurations without requiring intent-specific rules or predefined profiles for the sampled combinations. The ISR reaches 0.9 for some intents, indicating that all specified constraints are simultaneously satisfied in a substantial fraction of measurement windows. However, performance varies across intents and repetitions, showing that the ability to generate configurations for previously unseen requirements does not guarantee consistent constraint satisfaction.


\section{Conclusion}
This work introduced an \gls{llm}-driven xApp that translates changing network-level intents into traffic-control configurations at runtime in programmable \glspl{ran}, by combining \gls{llm}-based reasoning, asynchronous network-state snapshots, and deterministic validation. Experiments on an OpenAirInterface testbed show an average \gls{isr} of $0.91$ with Qwen3-8B, compared with $0.52$ for a static TC-RAN baseline. The experiments further demonstrate the controller's ability to adapt to reduced radio capacity, maintaining an \gls{isr} of $\approx 0.88$ under degraded channel conditions. 
Since configurations are selected at runtime, the controller accommodates new combinations of requirements without stopping the xApp, designing a dedicated configuration, and redeploying it. While implemented on TC-RAN, the proposed approach is applicable to other programmable traffic-control mechanisms that expose functions via the E2 interface.
Future work will address multi-UE scenarios with conflicting requirements, the fine-tuning of smaller \glspl{llm} to reduce inference latency, and memory techniques to reuse past decisions.

\balance
\bibliographystyle{IEEEtran}
\bibliography{biblio}

\end{document}